\documentstyle[pra,aps]{revtex}
\tighten
\begin{document}
\draft
\twocolumn
\title{Realization of a collective decoding of codeword states}
\author{Masahide Sasaki$^1$, Tsuyoshi Sasaki-Usuda$^{2}$, Masayuki
Izutsu$^1$, and Osamu Hirota$^{3}$}
\address{${}^{1}$Communications Research Laboratory, Ministry of Posts and
Telecommunications\\
 Koganei, Tokyo 184, Japan}
\address{${}^{2}$Nagoya Institute of Technology, 
 Gokiso-chou, Showa-ku, Nagoya 466, Japan}
\address{${}^{3}$Research Center for Quantum Communications, Tamagawa
University\\
 Tamagawa-gakuen, Machida, Tokyo 194, Japan}
 
\date{14 October 1997}

\maketitle

\begin{abstract}
A physical model for the optimum collective decoding that attains the
minimum average error probability in distinguishing codeword states is
presented. 
This model is based on a cavity QED technique which is available at
present.  It will open a possibility for a quantum decoder  that realizes
the superadditivity in classical capacity of quantum channel which was
demonstrated in the preceding paper. 
\end{abstract}



\section{Introduction}\label{sec1}

Distinguishing nonorthogonal quantum states at the minimum error is a
fundamental problem in quantum communication. The optimum strategy
minimizing the average error probability can be, in principle, derived from
a linear optimization in terms of Bayesian decision problem. Such a
strategy is generally represented by a probability operator measure (POM)
which is a set of nonnegative Hermitian operators $\hat\Pi_i$ satisfying
the resolution of the identity \cite{Helstrom_QDET}, 
\begin{equation}
\sum_i \hat\Pi_i=\hat I, \quad \hat\Pi_i\ge0.  
\end{equation}
Except for pure-state signals with a certain symmetry, it is tedious job to
derive explicit expressions for $\{\hat\Pi_i\}$ \cite{Osaki}. In addition,
even when they can be obtained, corresponding physical processes are not
necessarily obvious. Mathematically, the optimum POM can be specified in
the Hilbert space of the minimum dimension that a set of signal states
spans. However,  it can often hardly be interpreted physically.  For
practical physical implementation, the POM should be constructed in a
larger Hilbert space which can fully describe the physical system making
the signal states. 
Such examples have been known only in the certain cases of binary signals
\cite{Sasaki's}. 
In the preceding paper, the problem of decoding $M$-ary codeword states at
the minimum average error probability was discussed. For attaining the
minimum error, a {\it collective decoding} was essential in which each
codeword state is detected as a single state-vector rather than detected as
the individual letter states separately.  It was shown that when it is
applied to the properly selected codeword states, the {\it quantum gain} in
transmittable information can be obtained.

In this paper, we present a physical scheme for the optimum  collective
decoding of $M$-ary codeword states, particularly, in the case that the
letter states are pure and binary.   This scheme consists of a quantum
circuit and a simple separate measurement on the individual letter states.
In the quantum circuit, a certain unitary transformation is carried out on
the received  codeword state  and a superposition of the codeword states is
generated.  This unitary transformation is designed so that the minimum
average error probability is attained when the output from the circuit is
detected by the given separate measurement. 
In the following, the decoding scheme proposed in the preceding paper is
briefly reviewed in Sec. \ref{sec2}, and then quantum circuit structures
are presented in Sec. \ref{sec3}.  
In particular, a concrete model for distinguishing two codeword states of
length 2 at the minimum average error will be proposed for an experimental
demonstration of the principle of collective decoding. 
Sec. \ref{sec4} is for concluding remarks.

\section{Decoding scheme}\label{sec2}

Let  binary letter states be $\{ \vert+\rangle , \vert-\rangle\}$ whose
state overlap $\kappa=\langle+\vert-\rangle$ is assumed to be real.  By
concatenating them into block sequences of length $n$, the $2^n$ possible
sequences  are made. We then pick up $M$-ary sequences as codeword states
that are denoted as $\{  \vert S_1\rangle, \cdots,   \vert S_M\rangle   \}$
$(M\le2^n)$, and use them with input probabilities $\{ \zeta_1, \cdots,
\zeta_M \}$.  The rest of sequences are denoted as $\{  \vert
S_{M+1}\rangle, \cdots,   \vert S_{2^n}\rangle   \}$.  Since the codeword
states are linearly independent, they span the $M$-dim Hilbert space ${\cal
H}_s$.  Let the $n$-th extended Hilbert space be ${\cal H}_s^{\otimes n}$.
The optimum collective decoding is described by an  orthonormal set $\{ 
\vert\omega_1\rangle, \cdots,   \vert\omega_M\rangle   \}$ on ${\cal H}_s$.

This is achieved, in some cases,  by  the square-root measurement 
\cite{Helstrom_QDET,Holevo_SubOptMeas78,Hausladen_SubOptMeas95,Hausladen96_c
oding} given by, 
\begin{mathletters} 
\begin{eqnarray}
\vert\mu_i\rangle &\equiv& \hat\rho^{-{1\over2}} \vert\tilde S_i\rangle, \\
\hat\rho &\equiv& \sum_{i=1}^M \vert\tilde S_i\rangle\langle\tilde S_i\vert, \\
\vert\tilde S_i\rangle &\equiv& {\sqrt\zeta_i}\vert S_i\rangle.
\end{eqnarray}
\label{eqn:sqrt-mu}
\end{mathletters}

\noindent 
Let us define the Gram matrix ${\bf\Gamma}\equiv(\langle\tilde S_i\vert
\tilde S_j\rangle)$.  

\noindent 
{\bf Theorem}

{\it For $0\le\kappa<1$,  the square-root measurement
$\{\vert\mu_i\rangle\}$ becomes optimum when all of the diagonal components
of ${\bf\Gamma}^{1\over2}$ are equal. }

Unless it is not the case,  $\{\vert\mu_i\rangle\}$ forms, at least, an
orthonormal set in ${\cal H}_s$. 
 So it is connected  with the optimum measurement states
$\{\vert\omega_i\rangle\}$  via a unitary operator $\hat V$ in ${\cal H}_s$
as $\vert\omega_i\rangle=\hat V\vert\mu_i\rangle$.  A straightforward
method to construct such a unitary operator is the Bayes-cost-reduction
algorithm proposed by Helstrom \cite{Helstrom82}.  In this algorithm, one
chooses a pair of codeword states $\{ \vert  S_i\rangle , \vert S_j\rangle
\}$, solves the binary decision problem  in the plane  spanned by the
corresponding pair of measurement basis vectors 
$\{ \vert\mu_i\rangle , \vert\mu_j\rangle \}$,  
and get the revised basis vectors $\{ \vert\mu_i'\rangle ,
\vert\mu_j'\rangle \}$ which can be connected with the previous ones via a
$U(2)$-operator $V^{(1)}$ in ${\cal H}_s$ as,
\begin{mathletters} 
\begin{eqnarray}
\vert\mu_i'\rangle &=&V^{(1)} \vert\mu_i\rangle, \\
\vert\mu_j'\rangle &=&V^{(1)} \vert\mu_j\rangle.
\end{eqnarray}
\end{mathletters} 

\noindent 
After revising the basis vectors, the average error probability will
decrease or, at worst, remain the same. This kinds of steps are to be
continued till reaching the optimum point.   Thus starting from the
square-root measurement basis vectors, the optimum ones are derived as, 
\begin{equation}
\vert\omega_i\rangle =\cdots V^{(2)} V^{(1)} \vert\mu_i\rangle, \quad
\forall i.  
\end{equation}

In order to consider a physical scheme, we make an orthonormal set $\{ 
\vert\omega_1\rangle, \cdots,   \vert\omega_{2^n}\rangle   \}$ on the whole
space ${\cal H}_s^{\otimes n}$ by adding the other basis vectors obtained
by using the Schmidt orthogonalization, 
\begin{equation}
\vert\omega_i\rangle=
{
{\vert
S_i\rangle-\displaystyle\sum_{k=1}^{i-1}\vert\omega_k\rangle\langle\omega_k\
vert S_i\rangle}
\over
{\sqrt{1-\displaystyle\sum_{k=1}^{i-1}\vert\langle\omega_k\vert
S_i\rangle\vert^2}}  
 },  \quad (i=M+1,\cdots, 2^n), 
\label{eqn:orthogonal_set}
\end{equation}
We denote the expansion of all the sequences by the above basis vectors as,  
\begin{mathletters} 
\begin{equation}
\left(\begin{array}{c} \vert S_1\rangle \\
                                   \vdots \\
                                   \vert S_{2^n}\rangle    \end{array} \right)
=
{\bf B}
\left(\begin{array}{c} \vert \omega_1\rangle \\
                                   \vdots \\
                                   \vert \omega_{2^n}\rangle    \end{array}
\right).   
\end{equation}
\begin{equation}
{\bf B}=(B_{ij})=(\langle \omega_j\vert S_i\rangle).
\end{equation}
\end{mathletters} 

\noindent    
Let $\{ \vert a\rangle , \vert b\rangle\}$ be the measurement basis vectors
distinguishing the individual letter states. By concatenating them into the
block sequences of length $n$, the $2^n$ product basis vectors in ${\cal
H}_s^{\otimes n}$ are made. We pick up $M$-ary basis vectors from them, and
denote them as 
$\{  \vert A_1\rangle, \cdots,  \vert A_M\rangle \}$
, and the rest of them as 
$\{ \vert A_{M+1}\rangle, \cdots,    \vert A_{2^n}\rangle \}$. 
All the sequences can be expanded alternatively by these basis vectors as, 
\begin{mathletters} 
\begin{equation}
\left(\begin{array}{c} \vert S_1\rangle \\
                                   \vdots \\
                                   \vert S_{2^n}\rangle    \end{array} \right)
={\bf C}  \left(\begin{array}{c} \vert A_1\rangle \\
                                   \vdots \\
                                   \vert A_{2^n}\rangle    \end{array}
\right),   
\label{eqn:expansion_C} 
\end{equation}
\begin{equation}
{\bf C}=(C_{ij})=(\langle A_j\vert S_i\rangle).
\label{eqn:matrix_C}
\end{equation}
\end{mathletters} 

\noindent  
The two basis sets are connected via a unitary operator $\hat U$ on the
whole space ${\cal H}_s^{\otimes n}$ as,
\begin{mathletters}
\begin{equation}
\vert \omega_i\rangle =\hat U^\dagger \vert A_i\rangle, \quad (i=1,\cdots,2^n),
\end{equation}
where 
\begin{equation}
\hat U^\dagger \equiv \sum_{i,j}^{2^n} u_{ji} \vert A_j\rangle\langle A_i
\vert, \quad u_{ji}\equiv({\bf B}^{-1} {\bf C})_{ij}. 
\end{equation}
\label{eqn:def_U}
\end{mathletters}

\noindent  
Thus  the optimum {\it collective decoding} $\{\vert\omega_1\rangle,
\cdots, \vert\omega_M\rangle\}$ can be effected by (i) transforming the
codeword states $\{\vert S_i\rangle\}$ by the unitary transformation $\hat
U$, and (ii) applying the measurement $\{ \vert A_1\rangle, \cdots,  \vert
A_M\rangle \}$ into the transformed codeword states.  Note that $\{\vert
A_i\rangle\}$ are the product basis vectors hence represent the separate
measurement on the individual  letter states.  The unitary transformation
$\hat U$ plays a role of an {\it adaptor}  to this separate measurement
\cite{Sasaki's}. 
The minimum error probability is obtained as 
\begin{equation}
P_e({\rm opt})=1-\sum_{i=1}^M \zeta_i \vert\langle S_i\vert \hat
U^\dagger\vert A_i\rangle\vert^2. 
\end{equation}
When the codeword states have a certain symmetry, the construction of $\hat
U$ will become much easier by choosing  the measurement basis vectors $\{
\vert A_1\rangle, \cdots,  \vert A_M\rangle \}$ taking that symmetry into
account.  See the example later.

Construction  of  the quantum circuit for $\hat U$  can be done i) by
decomposing it into $U(2)$-operators $\hat T_{[j,i]}$ by applying the
algorithm proposed by Reck and others \cite{Reck94} as, 
\begin{mathletters}
\begin{equation}
\hat U =\hat T_{[2,1]}  \hat T_{[3,1]} \cdots \hat T_{[2^n,2^n-2]} \hat
T_{[2^n,2^n-1]}, \label{eqn:decomp_a} 
\end{equation}
where 
\begin{equation}
\hat T_{[j,i]} ={\rm exp}[-\gamma _{ji}(\vert A_i\rangle\langle A_j\vert -
\vert A_j\rangle\langle A_i\vert)]       ,
\label{eqn:decomp_b}
\end{equation}
\end{mathletters}

\noindent 
and ii) by applying the formula established by Barenco. et. al. for
simulating a discrete unitary operator \cite{Barenco95}.  
Here the following  point should be noted. Since qubits in the circuit are
the letter states themselves constituting the codeword states, the gates
should consist of the single physical species from which the letter states
are made. Such gates known so far are Sleator and Weinfurter's gate
consisting of two-state atoms \cite{Sleator95} and the quantum phase gate
acting on two photon-polarization states \cite{Turchette95}.

\section{Quantum circuit structures}\label{sec3}

In this section, the structure of the quantum circuit is presented. A model
using  a physically two-state system such as a two-level atom or a pair of
single-mode photon polarizations is considered first, aiming at an
experimental demonstration of the collective decoding. Then, an
implementation in the case of coherent-state  signals is discussed.

\subsection{Physically two-state system}\label{sec3-1}

Let $\{ \vert\uparrow\rangle , \vert\downarrow\rangle \}$ be the
upper-level and lower-level states of an atom, or two orthogonal linear
polarization states of a single-mode optical field. The measurement basis
vectors $\{ \vert a\rangle , \vert b\rangle \}$ can then be taken naturally
as $\{ \vert\uparrow\rangle , \vert\downarrow\rangle \}$ representing a
level detection or a linear polarizer. 
Suppose that the letter states are made by rotating the state
$\vert\uparrow\rangle$ by the angle $\theta$ or  $\pi - \theta$ around the
y-axis, 
\begin{mathletters}
\begin{eqnarray}
\vert+\rangle&=&\hat
R_y(\theta)\vert\uparrow\rangle={\sqrt{1-p}}\vert\uparrow\rangle -
{\sqrt{p}}\vert\downarrow\rangle, 
\\
\vert-\rangle&=&\hat
R_y(\pi-\theta)\vert\uparrow\rangle={\sqrt{p}}\vert\uparrow\rangle -
{\sqrt{1-p}}\vert\downarrow\rangle, 
\end{eqnarray}
\end{mathletters}

\noindent 
where
\begin{equation}
\hat R_y(\theta)=\left(  \begin{array}{cc}
                                                {\rm cos}{\theta\over2} &
{\rm sin}{\theta\over2}      \cr
                                              -{\rm sin}{\theta\over2}  &
{\rm cos}{\theta\over2}           \end{array}   \right)
=\left(  \begin{array}{cc}
                                                {\sqrt{1-p}} & {\sqrt{p}}  
   \cr
                                              -{\sqrt{p}}  & {\sqrt{1-p}}  
        \end{array}   \right), 
\end{equation}
and $p=(1 - {\sqrt{1-\kappa^2}} )/2$.

Let us consider a simple case  of distinguishing two codeword states $\{
\vert++\rangle , \vert--\rangle \}$, where $\vert++\rangle$ means 
$\vert+\rangle\otimes\vert+\rangle$, etc.  
The two measurement basis vectors of the optimum collective decoding are
given by, 
\begin{mathletters}
\begin{eqnarray}
\vert\omega_1\rangle&=&
       \sqrt {{1-p_2}\over{1-\kappa^4}} \vert++\rangle
      - \sqrt {{p_2}\over{1-\kappa^4}} \vert--\rangle, \\
\vert\omega_2\rangle&=&
        - \sqrt {{p_2}\over{1-\kappa^4}} \vert++\rangle
        + \sqrt {{1-p_2}\over{1-\kappa^4}}\vert--\rangle, 
\end{eqnarray}
\end{mathletters}

\noindent 
where $p_2=(1 - {\sqrt{1-\kappa^4}} )/2$ is the minimum bound of the
average error probability. 
The codeword states are expanded as, 
\begin{mathletters}
\begin{eqnarray}
\vert++\rangle&=&(1-p)\vert\uparrow\uparrow\rangle \nonumber \\ 
                        &-& {\sqrt{(1-p)p}}(\vert\uparrow\downarrow\rangle
+ \vert\downarrow\uparrow\rangle)
                         +p\vert\downarrow\downarrow\rangle , 
\\
\vert--\rangle&=&p\vert\uparrow\uparrow\rangle  \nonumber \\
                         &-& {\sqrt{(1-p)p}}(\vert\uparrow\downarrow\rangle
+ \vert\downarrow\uparrow\rangle)
                         +(1-p)\vert\downarrow\downarrow\rangle,  
\end{eqnarray}
\label{eqn:expan_codeword}
\end{mathletters}

\noindent  
here again
$\vert\uparrow\uparrow\rangle=\vert\uparrow\rangle\otimes\vert\uparrow\rangle$, etc.
Let us denote 
$\vert\uparrow\uparrow\rangle$, $\vert\uparrow\downarrow\rangle$,
$\vert\downarrow\uparrow\rangle$ and $\vert\downarrow\downarrow\rangle$ 
as 
$\vert A_1\rangle$, $\vert A_2\rangle$, $\vert A_3\rangle$ and $\vert
A_4\rangle$, respectively.  
It is easy to see the optimum measurement basis vectors can be expressed as, 
\begin{mathletters}
\begin{eqnarray}
\vert\omega_1\rangle&=&\hat U^\dagger \vert A_1\rangle, \\
\vert\omega_2\rangle&=&\hat U^\dagger \vert A_4\rangle, 
\end{eqnarray}
\end{mathletters}

\noindent    
where the unitary operator $\hat U$ is defined by the following matrix
representation in terms of the basis vectors 
$\{ \vert A_1\rangle,\vert A_2\rangle,\vert A_3\rangle,\vert A_4\rangle \}$, 
\begin{equation}
\hat U={1\over{2d_0}}
           \left(  \begin{array}{cccc}
                        1+d_0  & -\kappa  & -\kappa & 1-d_0    \cr
                        \kappa & 1+d_0  & 1-d_0 & \kappa  \cr
                        \kappa & 1-d_0  & 1+d_0 & \kappa  \cr 
                        1-d_0  & -\kappa  & -\kappa & 1+d_0           
                      \end{array}   \right), 
\end{equation}
with $d_0=\sqrt{1+\kappa^2}$.  Thus the optimum collective decoding by
$\{\vert\omega_1\rangle, \vert\omega_2\rangle\}$ has been decomposed into
the unitary transformation by $\hat U$ and the separate measurement by 
$\{\vert A_1\rangle,\vert A_4\rangle\}$.   As seen in Eq.
(\ref{eqn:expan_codeword}),  the codeword states $\vert++\rangle$ and 
$\vert--\rangle$  mainly consists of $\vert\uparrow\uparrow\rangle$ and
$\vert\downarrow\downarrow\rangle$, respectively, when  the letter-state
overlap $\kappa$ is small. In addition, the expansion of   Eq.
(\ref{eqn:expan_codeword}) is symmetric in terms of these basis vectors.
Therefore, 
choosing $\{\vert A_1\rangle,\vert A_4\rangle\}$ as the final measurement
seems very natural, and actually simplifies a construction of  a quantum
circuit for $\hat U$.

The  transformed codewords $\{\hat U\vert++\rangle, \hat U\vert--\rangle\}$
are always within the space spanned by $\{\vert A_1\rangle,\vert
A_4\rangle\}$. The output is always either `$\uparrow\uparrow$' or
`$\downarrow\downarrow$'.   
The decomposition of the unitary operator $\hat U$ into the
$U(2)$-operators can be done in the following way, 
\begin{equation}
\hat U =\hat T_{[2,1]}  \hat T_{[3,1]} \hat T_{[3,2]} \hat T_{[4,1]} \hat
T_{[4,1]} \hat T_{[4,2]} \hat T_{[4,3]},  
\label{eqn:U_decomp}
\end{equation}
where the rotation angles $\gamma _{ji}$ in $\hat T_{[j,i]}$ are determined by, 
\begin{mathletters}
\begin{equation}
{\rm cos}{\gamma _{43}\over2}={{d_0 + 1}\over{d_1}}, \quad {\rm sin}{\gamma
_{43}\over2}=-{{\kappa}\over{d_1}}, 
\end{equation}
\begin{equation}
d_1=\sqrt{(d_0+1)^2 + \kappa^2}, 
\end{equation}
\begin{equation}
{\rm cos}{\gamma _{42}\over2}={{d_1}\over{d_2}}, \quad {\rm sin}{\gamma
_{42}\over2}=-{{\kappa}\over{d_2}}, 
\end{equation}
\begin{equation}
d_2=\sqrt{d_1^2 + \kappa^2},
\end{equation}
\begin{equation}
{\rm cos}{\gamma _{41}\over2}={{d_2}\over{d_1}}, \quad {\rm sin}{\gamma
_{41}\over2}=-{{d_0 - 1}\over{2d_0}}, 
\end{equation}
\begin{equation}
\gamma _{32}=\gamma _{41}, \quad  \gamma _{31}=-\gamma _{42}, \quad \gamma
_{21}=-\gamma _{43}. 
\end{equation}
\end{mathletters}

\noindent    
The above $U(2)$-rotations $\hat T_{[j,i]}$'s can be performed by the
quantum circuits shown in Fig. \ref{fig1}.   All the notations that are not
explained particularly are  borrowed from ref. \cite{Barenco95}.  In the
figure, the time evolves from the left to the right. Denoting a quantum
state to be processed as $\vert\mu\nu\rangle$
=($\vert\mu\rangle\otimes\vert\nu\rangle$),  the upper and lower lines
correspond to the evolutions of the first (with the initial state
$\vert\mu\rangle$) and second (with the initial state $\vert\nu\rangle$)
qubits, respectively.  The 2-bit gates in these circuits, $\bigwedge_1(\hat
R_y(\gamma_{ji}))$ can be further decomposed into the circuits of the type
shown in Fig. \ref{fig2}.  Thus  $\hat T_{[j,i]}$'s  can be carried out by
the circuits consisting of the 1-bit gates and the controlled-NOT gates.

As mentioned in the previous section, the candidates to implement these
gates are 
Sleator and Weinfurter's gate  for two-state atoms \cite{Sleator95} and the
quantum phase gate for  binary photon polarizations \cite{Turchette95}. 
For readers' convenience, we give explicit constructions of the gates
required in our circuits, supplementing the original paper \cite{Sleator95}
by practical formula for our particular application.  As shown later for
the case $n=3$, basic gates for our purpose are the 2-bit gate
$\bigwedge_1(\sqrt{\sigma_x})$, the 1-bit gates $\bigwedge_0(\sigma_x)$ and
$\bigwedge_0(\hat R_y(\gamma))$.  The controlled-NOT gate
$\bigwedge_1(\sigma_x)$ is obviously equivalent to
$\bigwedge_1(\sqrt{\sigma_x})\bigwedge_1(\sqrt{\sigma_x})$.

Implementations of the 1-bit gates are straightforward by using the Ramsey
zone (RZ) which is characterized by the Hamiltonian, 
\begin{eqnarray}
\hat H_{RZ}&=&{1\over 2} \hbar\nu (
\vert\uparrow\rangle\langle\uparrow\vert -
\vert\downarrow\rangle\langle\downarrow\vert ) \nonumber\\
&+& i \hbar \vert\epsilon\vert ( {\rm e}^{-i\nu
t}\vert\uparrow\rangle\langle\downarrow\vert - {\rm e}^{i\nu
t}\vert\downarrow\rangle\langle\uparrow\vert) , 
\end{eqnarray}
whose time evolution is described by  the unitary operator, 
\begin{equation}
\hat U_R (\tau,\vert\epsilon\vert)=
\left(
\begin{array}{cc}
{\rm e}^{-i\nu\tau/2}{\rm cos}(\vert\epsilon\vert\tau) & {\rm
e}^{-i\nu\tau/2}{\rm sin}(\vert\epsilon\vert\tau)      \cr
 -{\rm e}^{i\nu\tau/2}{\rm sin}(\vert\epsilon\vert\tau) & {\rm
e}^{i\nu\tau/2}{\rm cos}(\vert\epsilon\vert\tau)
\end{array}
\right)
\end{equation}
in the spinor representation, where  $\vert\epsilon\vert$ is the pumping
field amplitude, the angular frequency $\nu$ corresponds to an atomic level
separation, and  $\tau$ is an interaction period.  This is capable of
implementing the following 1-bit operations: 
\begin{eqnarray}
\hat R_y(\theta)&=&\left(  \begin{array}{cc}
                                                {\rm cos}{\theta\over2} &
{\rm sin}{\theta\over2}      \cr
                                              -{\rm sin}{\theta\over2}  &
{\rm cos}{\theta\over2}           \end{array}   \right), 
\\
\hat R_z(\theta)&=&\left(  \begin{array}{ll}
                                                {\rm exp}(i\theta/2) & 0   
  \cr
                                                 0  & {\rm exp}(-i\theta/2)
          \end{array}   \right). 
\end{eqnarray}
By using them, $\hat R_x(\theta)$ can  be realized as, 
\begin{equation}
\hat R_x(\theta)=\left(  \begin{array}{cc}
                                                {\rm cos}{\theta\over2} &
i{\rm sin}{\theta\over2}      \cr
                                              i{\rm sin}{\theta\over2}  &
{\rm cos}{\theta\over2}          \end{array}   \right) 
                          =\hat R_z({\pi\over2})\hat R_y(\theta)\hat
R_z(-{\pi\over2}).
\end{equation}
$\hat R_x(\pi)$ plays a role of $\bigwedge_0(\sigma_x)$.  We also introduce
the other rotations for later purpose, 
\begin{eqnarray}
\hat U_{R1}&=&\hat U_R (\tau,\vert\epsilon\vert) \mbox{ with }
\vert\epsilon\vert\tau={\pi\over4}, \\
\hat U_{R2}&=&\hat U_R (\tau',\vert\epsilon'\vert) \mbox{ with }
\vert\epsilon'\vert\tau'={\pi\over4}, 
\end{eqnarray}
where $\tau\ne\tau'$ in general.

On the other hand, the implementation of the 2-bit gate employs  a
microcavity and  the Ramsey zones. The atom-cavity field interaction is
described by the Jaynes-Cummings Hamiltonian, 
\begin{eqnarray}
\hat H=\hbar \omega {\hat a}^\dagger {\hat a} 
&+& {1\over 2} \hbar\nu ( \vert\uparrow\rangle\langle\uparrow\vert -
\vert\downarrow\rangle\langle\downarrow\vert )  \nonumber  \\
&+& \hbar g ( {\hat a}^\dagger\vert\downarrow\rangle\langle\uparrow\vert 
                    + {\hat a}\vert\uparrow\rangle\langle\downarrow\vert ), 
\end{eqnarray}
where ${\hat a}$ (${\hat a}^\dagger$) is an annihilation (creation)
operator for the cavity field with the angular frequency $\omega$, and $g$
is the coupling constant between the cavity  field and the atom.  
It is assumed that $\nu$ is originally detuned from the cavity resonant
frequency $\omega$ so that the atom undergoes an off-resonant interaction
whose time evolution is given as, 
\begin{eqnarray}
\hat U_{\rm off}(t)&=&\sum_{n=0}^\infty \vert n\rangle\langle n\vert 
\nonumber \\
&\otimes&
\left(
\begin{array}{cc}
{\rm e}^{-i({\nu\over2}+g_{\rm eff})t-in g_{\rm eff}t} & 0     \cr
 0 & {\rm e}^{{{i\nu t}\over2}+in g_{\rm eff}t}
\end{array}
\right),
\end{eqnarray}
where $g_{\rm eff}=g^2/\delta$, $\delta=\nu-\omega$, and $\vert n\rangle$
is $n$-photon state. Phase factors involving $\omega$ have been  omitted
since it will give no physical effect. If $\nu$ is tuned to $\omega$ by an
appropriate Stark shifting, an on-resonant interaction can be carried out
as, 
\begin{equation}
\hat U_{\rm on}=
\left(
\begin{array}{cc}
0 & -i  \vert 0\rangle\langle 1\vert    \cr
-i  \vert 1\rangle\langle 0\vert &  \vert 0\rangle\langle 0\vert
\end{array}
\right),
\end{equation}
where the interaction period $t_0$ is chosen as $gt_0={\pi\over2}$ and the
fact is taken into account that the cavity field is either $\vert 0\rangle$
or $\vert 1\rangle$ throughout the gate operation.  Denoting the control-,
target-bit atoms and the cavity as `a' ,`b' and `c', respectively,
$\bigwedge_1(\sqrt{\sigma_x})$ can be realized by applying a unitary
process, 
\begin{eqnarray}
\hat R_z^{(a)}(-{5\over4}\pi&)&\hat R_x^{(a)}(\pi)\hat U_{\rm on}^{(a,c)}
\hat U_R^{(b)} (\tau',\vert\epsilon'\vert)  \nonumber\\
&\cdot&  \hat U_{\rm off}^{(b,c)}(t) \hat U_R^{(b)} (\tau,\vert\epsilon\vert)
\hat U_{\rm on}^{(a,c)} \hat R_x^{(a)}(\pi)
\label{eqn:uuu}
\end{eqnarray}
where the superscript indicates on what system(s) the operator acts. Here
$\vert\epsilon\vert\tau=\vert\epsilon'\vert\tau'={\pi\over4}$  and 
$$
i{\nu(\tau-\tau')\over2}-i{\nu t\over2} - i {g_{\rm eff}t\over2} = 2\pi n
\quad (n={\rm integer}), 
$$
should be satisfied.

In the decoder based on the above cavity QED system,   either of the
codeword states $\vert++\rangle$ or $\vert--\rangle$ passes through the
sequence of the Ramsey zones and the cavities that are mounted according to
the circuit for Eq. (\ref{eqn:U_decomp}), and then detected by a level
detector.

In the case of binary photon polarizations which is more practical system
for communication,  the decoder structure is quite parallel to the
two-state atomic case by replacing the  Ramsey zone and the 2-bit gate with
a polarizer and the quantum phase gate, respectively.

So far,  distinguishing codeword states at the minimum average error
probability has never been done even for the simplest case $\{
\vert++\rangle , \vert--\rangle \}$. 
Possible methods for it are not necessarily the above kind of scheme.  
As shown by Brody and Meister \cite{Brody96}, the separate measurement
together with the suitable feedback arrangement is capable of
distinguishing the {\it two} codeword states at the minimum average error
probability. For codeword states made of spin particles, the generalized
Stern-Gerlach measurement may effect the optimum collective decoding
\cite{Swift80}.  
As yet, however,  it is strongly desired to demonstrate the scheme based on
quantum circuits plus a simple standard measurement proposed in this paper
because it seems the most natural and systematic method for designing a
practical decoder.

Recently,  the first experiment of distinguishing binary photon
polarizations $\{ \vert+\rangle , \vert-\rangle \}$ at the minimum average
error probability $p$ was done by Barnett and Riis \cite{Barnett97_exp}. 
It has a significant meaning for optical communication.  The next step
might be to demonstrate the discrimination of $\{ \vert++\rangle ,
\vert--\rangle \}$ at the error rate $p_2$. It will be possible if the
Barnett-Riis experiment is assisted with the circuits involving the quantum
phase gate.  Once a breakthrough is done in this direction, an extension to
the optimum collective decoding for $M$-ary  codeword states of optical
polarizations might be straightforward.

From an information theoretic point of view, the optimum collective
decoding of the four codeword states of length $n=3$ is of significant
importance, because the superadditivity of classical capacity can appear as
shown in the preceding paper. The circuits for $\hat T_{[ji]}$'s becomes
more complicated. It might be worth giving one of such circuits.  The
principle of the formula can easily be understood by showing an example,
say, a rotation $\hat T_{6,3}={\rm
exp}[-\gamma(\vert\uparrow\downarrow\uparrow\rangle\langle\downarrow\uparrow
\downarrow\vert -
\vert\downarrow\uparrow\downarrow\rangle\langle\uparrow\downarrow\uparrow\vert)]$. It can be executed by the circuit shown in Fig. \ref{fig3}. 
The block denoted as $\hat M$ is for mapping $\{
\vert\uparrow\downarrow\uparrow\rangle ,
\vert\downarrow\uparrow\downarrow\rangle \}$ into 
$\{ \vert\downarrow\downarrow\uparrow\rangle ,
\vert\downarrow\downarrow\downarrow\rangle \}$.  In the mapped plane, the
desired rotation is carried out as the 3-bit gate operation 
$\bigwedge_2(\hat R_y(2\gamma))$. The two 3-bit gates in Fig. \ref{fig3}
can be further decomposed into circuits of the 2-bit gates as illustrated
in Fig. \ref{fig4}.  
Thus it is easy to see that the basic gates for our circuits are the 1-bit
gates 
$\bigwedge_0(\hat R_y(\pm\gamma))$ and $\bigwedge_0({\sigma_x})$, and the
2-bit gate $\bigwedge_1(\sqrt{\sigma_x})$.

\subsection{Binary phase-shift-keyed signals}\label{sec3-2}

We would like to mention the case of binary phase-shift-keyed (BPSK)
signals of optical coherent states $\{ \vert\alpha\rangle,
\vert-\alpha\rangle \}$. 
This signal system is the most basic keying in long distant and ultra fast
coherent light communications. 
If the collective decoding for this case could be  realized,  it will give
a great impact on communication technology.  
Although the letter states  are binary,  the signal system is {\it not} a
physically {\it two-state} system in this case. 
Therefore in order to apply the above decoding scheme,  a new class of
gates  must be invented.   
Their specifications depend directly on what kind of measurement basis
vectors are chosen as $\{\vert a\rangle,\vert b\rangle\}$.

Let us consider the following example. 
We first transform the received codeword state by combining each field  of
the letter state, $\vert\alpha\rangle$ or $\vert-\alpha\rangle$,
sequentially  with a local oscillating field with very large intensity via
a beam splitter having almost perfect transmittance. Each process is
represented by a unitary operator 
$\hat D(\alpha)={\rm exp}(\alpha\hat a^\dagger - \alpha^\ast\hat a)$. The
transformed  codeword states consist of the new letter states
$\{\vert0\rangle,\vert-2\alpha\rangle\}$.   The optimum collective decoding
is then performed for the  transformed  codeword states.    
In that decoding they undergo further the certain unitary transformation
$\hat U$ and are detected by the separate measurement  that distinguishes
each letter state 
$\vert0\rangle$ or $\vert-2\alpha\rangle$ according to the measurement
basis vectors $\vert a\rangle$ or $\vert b\rangle$ such as 
\begin{mathletters}
\begin{eqnarray}
\vert a\rangle&=&\vert0\rangle, \\
\vert
b\rangle&=&{{\vert-2\alpha\rangle-\vert0\rangle\langle0\vert-2\alpha\rangle}
                                                     \over
\sqrt{1-\vert\langle0\vert-2\alpha\rangle\vert^2}}. 
\end{eqnarray}
\end{mathletters}

\noindent 
The state $\vert b\rangle$ only includes the Fock states with finite
photons.  Therefore this final measurement 
$\{ \vert a\rangle\langle a\vert  , \vert b\rangle\langle b\vert \}$ is
equivalently accomplished by the photon counting
$\{\vert0\rangle\langle0\vert  ,  \displaystyle\sum_{n=1}^\infty \vert
n\rangle\langle n\vert  \}$.

The quantum circuit realizing the required unitary transformation $\hat U$
should act on the qubits consisting of $\{ \vert 0\rangle, \vert
b\rangle\}$.  
In particular, the basic 2-bit gate $\bigwedge_1(\sqrt{\sigma_x})$  is
specified as,  
\begin{eqnarray}
\vert0\rangle_a \vert0\rangle_b \vert\psi\rangle_c  
           &\rightarrow&  \vert0\rangle_a \vert0\rangle_b
\vert\psi\rangle_c \nonumber\\
\vert0\rangle_a \vert b\rangle_b \vert\psi\rangle_c  
           &\rightarrow&  \vert0\rangle_a \vert b\rangle_a 
\vert\psi\rangle_c \nonumber\\  
\vert b\rangle_a \vert0\rangle_b \vert\psi\rangle_c  
           &\rightarrow& \vert b\rangle_a  
           {1\over{\sqrt2}}({\rm e}^{i\pi/4} \vert0\rangle_b + {\rm
e}^{-i\pi/4} \vert b\rangle_b) \vert\psi\rangle_c \nonumber\\
\vert b\rangle_a \vert b\rangle_b  \vert\psi\rangle_c  
           &\rightarrow&  \vert b\rangle_a 
           {1\over{\sqrt2}}( {\rm e}^{-i\pi/4}\vert0\rangle_b + {\rm
e}^{i\pi/4}\vert b\rangle_b) \vert\psi\rangle_c,  \nonumber
\end{eqnarray}
where $\vert\psi\rangle_c$ represents a certain ancillary system. 
That is, depending on whether the control bit is the vacuum state or
consists only of the finite-photon Fock states, the target bit should
remain unchanged or should be transformed into the superposition between
$\vert0\rangle$ and $\vert b\rangle$, respectively. 
It is an open question to find physical models accomplishing this function.

\section{Concluding remarks}\label{sec4}

As shown in the preceding paper, the collective decoding plays an essential
role to realize true benefits of quantum communication. 
A systematic design of such a decoder becomes  possible when techniques
established in quantum computation are applied.  
The concrete scheme for the simplest collective decoding of $\{
\vert++\rangle , \vert--\rangle \}$ at the minimum error rate was
presented. 
It is implementable by use of the current cavity QED technique. 
Experimental demonstration is strongly desired.  Further it is a
challenging task to demonstrate experimentally the information theoretic
quantum gain $I_n/n-C_1 > 0$ which must be seen for the codeword states $\{
\vert+++\rangle,\vert+--\rangle,\vert--+\rangle,\vert-+-\rangle  \} $.  The
amount of gain is quantitatively small but it will become measurable by
increasing numbers of trial of transmission.   
From a practical view point of coherent light communication, the problem
was addressed on an implementation of a new class of gates that can act on
the binary coherent states.

\acknowledgments

The authors would like to thank  Dr. H. Mabuchi of California Institute of
technology,  Dr. M. Ban of Hitachi Advanced Research Laboratory, Dr. K.
Yamazaki and Dr. M. Osaki of Tamagawa University, Tokyo, for their helpful
discussions.

\begin{figure}
\caption{ The diagram representing the quantum circuits for realizing the
$U(2)$-rotations in Eq. (17).  The time evolves from the left to the right.
Denoting a quantum state to be processed as $\vert\mu\nu\rangle$
($=\vert\mu\rangle\otimes\vert\nu\rangle$),  the upper and lower lines
correspond to the evolutions of the first (with the initial state
$\vert\mu\rangle$) and second (with the initial state $\vert\nu\rangle$)
qubit, respectively.}
\label{fig1}
\end{figure}

\begin{figure}
\caption{ The diagramatic  representation for a decomposition of the
rotation of one qubit around the $y$-axis $\hat R_y(\gamma)$ conditioned by
the state of the other qubit.}
\label{fig2}
\end{figure}

\begin{figure}
\caption{ The quantum circuit effecting the rotation $\hat T_{6,3}={\rm
exp}[-\gamma(\vert\uparrow\downarrow\uparrow\rangle\langle\downarrow\uparrow
\downarrow\vert -
\vert\downarrow\uparrow\downarrow\rangle\langle\uparrow\downarrow\uparrow\vert)]$ which is used for constructing the decoder of the codeword states of length 3. }
\label{fig3}
\end{figure}

\begin{figure}
\caption{ The diagramatic  representation for decompositions of the 3-bit
gates $\bigwedge_2(\sigma_x)$ and $\bigwedge_2(\hat R_y(2\gamma))$ into the
2-bit gates.}
\label{fig4}
\end{figure}

\end{document}